# Math Marvel with M-Learning

Dr.RD.Balaji[#1], Dr.V.Veeramani[#2], Er.Malathi Balaji[*3]
[#] College of Applied Sciences- Salalah, Sultanate of Oman
[1] balajicas@yahoo.com
[2] veeramani.sal@cas.edu.om

[*] Madurai Kamaraj University, India
[3] bmalathisai@gmail.com

***Abstract*** - **Math is the backbone of any field. Still it's a night mare for many. Recent survey proves that many students become dropouts from their higher education due to math courses. ICT is an enchanted word in the contemporary educational environment. It made the learning process more entertaining and almost made the "knowledge loss" negligible. Adopting the ICT in math courses are still in the infant level. Hence it's a challenge placed in front of the IT and academic professionals teaching math to make a suitable ICT tools for math courses to make the learning an amusing experience. In this paper we have highlighted three main concepts which make the math classes in a fascinating way. The first method is introducing revolutionary hybrid e-books which make the reading with both audio and video facilities. The second method is facilitating the flip class room so that student may have anywhere-anytime learning experience. Finally the recent trends in m-learning are using apps for math courses. We have highlighted the improvement showed by the students based on the survey conducted with two groups of students with m-learning tools and without m-learning tools. Even though there are good improvements showed by the students, we the researchers feel that few more improvements are required in these methodologies. The suggestions for the same are made in the recommendations and conclusion section.**

## I. INTRODUCTION

Education is defined in a different way by dictionaries and academicians. But in broader sense it is viewed in two different perspectives. They are teaching and learning. The learning is nothing but any action or continuous actions performed by a person which make some formative effects on the mind of that person. Mainly education helps to transmit the accumulated knowledge, skills, ethics, values and culture of the society to the next generation deliberately [1].

The traditional teaching differs a lot from the contemporary teaching due to the influence of the contemporary technologies. The technology education still differs from the normal learning process because it needs to learn not only the knowledge but also skills in that technology. Technology education needs the knowledge and skills from the components of science, technology, engineering and mathematics (STEM) [2]. In general, technology helps humankind to make their life easy. We do most of our jobs fast and easy because of the advancements in the technology. All technologies are having application of tools, materials, processes and systems done by us to solve our problems. Apart from the normal learning tools like reading books, listening and viewing videos and audio files, the technology education needs to have practice of few skills by activities in the labs [3]. In the technology learning knowledge of content, processes and skills should be used together to effectively engage students and promote a complete understanding of the sciences, related technologies and their interrelationship. This can be done with the help of mathematics courses in that discipline.

Nowadays, in all countries, technological education is mainly focused to show their strength and development to other countries. Hence all universities made the math courses compulsory in all disciplines. Since students in higher education are facing problem in understanding math and its applications in their field, we could see many dropouts or students with less CGPA during their higher education degree. Hence Instructional technology is another stream which mostly concentrate on the delivering the courses to students through ICT. In this paper we are concentrating on delivering the mathematical courses with the help of instructional technology to the higher education students.

## II. E/M LEARNING

"*Any type of learning that takes place in learning environments and spaces that take account of the mobility of technology, mobility of learners and mobility of learning*" [4]. The above definition is given for m-learning and it is mostly accepted by many of the researchers. M-learning is also viewed as a component of e-learning. M-learning started emerging after 2004 rapidly because of the advancements in the mobile phone devices and tabs introduction. Like any other technologies m-learning also was having lots of issues in the infant stage and by now it has overcome most of the disadvantages, as well as new facilities also provided to the m-learners which were not proposed even by researchers in the initial stages of m-learning [5]. The Cloud concepts in the IT field influenced most of the m-learning strategies in providing common materials in a most economical way internationally with security. Many researchers have compared the m-learning with e-learning and agreed that m-learning is the extension of e-learning and e-learning with anywhere-anytime facility becomes m-learning [6]. In a recent survey with the higher education students in Oman, proved that 100 percent of the students are having smart phone and they are well equipped with the basic infrastructure required for implementing the m-learning. Hence we have decided to start doing our research with the m-learning environment and also provided options to the students to go with e-learning. In the recent mobile phones we are having more interactive options to make communication with a group of people easily [7]. So we have decided to go for m-learning environment than e-learning because of the above mentioned advantages. The math courses demand more practice and time from the students. Hence m-learning will be the proper tool for the students to get in touch with the math course materials Just-In-Time (JIT) and Anywhere-Anytime (AWAT).

## III. HYBRID E-BOOKS

"*An electronic book (variously: e-book, eBook, e-Book, ebook, digital book, or even e-edition) is a book-length publication in digital form, consisting of text, images, or both, readable on computers or other electronic devices*"[8]. Even though there are some arguments that the e-books will not give the feeling for reading an actual book, there are so many advantages are also there in the e-books. The recent e-books are having facility of keeping the audio and video files in it. The recent surveys proved that people are spending more time with their electronic gadgets [15]. Instead of seeing this as a negative trend, we can make good out of this trend through e-books. Preparing e-books by the academicians is not difficult like a decade before. There are so many e-book creation software's available in the market for either nominal cost or for free. Each and every tool has its own advantages and disadvantages. The comparison of these tools is showed in table 1. We can expect more options from these tools in near future.

TABLE 1
E-BOOK CREATION SOFTWARE COMPARISON (SOURCE: HTTP://EN.WIKIPEDIA.ORG/WIKI/LIST_OF_E-BOOK_SOFTWARE) [16]

| **Brand Name / Feature** | **Open Source** | **Page Flip** | **e-Archive** | **Audio/ Video** | **Software Installation** | **Publish/ Share Online** | **Work Online** | **Work Offline** | **For PC** | **For Mac** |
|---|---|---|---|---|---|---|---|---|---|---|
| **flipb Software** | No | Yes | Yes | Yes | Yes | Yes | Yes | Yes | Yes | Yes |
| **ePaperFlip** | No | Yes | Yes | Yes | No | Yes | Yes | No | Yes | Yes |
| **PUB HTML5** | No | Yes | Yes | Yes | Yes | Yes | Yes | Yes | Yes | Yes |
| **kvisoft flipbook maker pro** | No | Yes | Yes | Yes | Yes | Yes | Yes | Yes | Yes | Yes |

## IV. FLIPPED CLASSROOM

Flipped Classrooms are a boon for academicians, which helps to utilize the students time even after their class hour for teaching using various technologies like QR code or E-books [9][10]. " The flipped classroom inverts traditional teaching methods, delivering instruction online outside of class and moving "homework" in to the class room" [11]. Flipped class room helps the slow learners to learn their courses at their own pace. Contemporary technologies help learners to communicate with their peer or their lecturers with the help of social network, forums, live chat, etc.,. The face to face interaction facility in the traditional teaching is replaced by these technologies and make the students learn comfortably even outside the classroom. In the flipped classroom the main concept used is the activity based learning [11]. Since 2007 this methodology used by the academicians and now it is also used by the Massive Open Online Courses (MOOCs). Many researchers proved that the flipped classroom helps to improve the understanding of the students in a specific topic and give the better result in overall [9] [10]. Flipped classrooms are very much helpful to the students to revisit the lectures, the lectures being small in size (8 to 15 minutes), when they don't understand the concepts. We are trying to use the same concept in the College of Applied Sciences Salalah for the network courses so that students will get more time for the lab practices and also they don't lose time for the conceptual topics.

## V. MATH APPS

One smart phone is enough to bring the complete world to our palm. Whether you have iPhone or android smart phone if you go to Google play or the app store of that mobile company, millions of apps are available at free of cost or at minimal cost related to any domain which we are interested with. The faculties of College of Applied Sciences-Salalah (CAS-S) were looking for a suitable app to deliver discrete mathematics we have found lots of apps to learn this course. By the rating, number of downloads and the comments of the user, we have selected an app developed by Wolfram Group named as "Discrete Math Course Assistant" [12]. The cost of this app was very nominal (4.9 $). It supports the students to learn as well as to get assistance in solving problems related to all topics covered in discrete mathematics course in CAS-S.

We were also giving other apps information for all levels of students to learn math in easy and facetious way. There are millions of apps available for leaning math. Hence we have recommended the apps which are highly preferred by many worldwide [13]. We have also listed few math apps which are popular among smart phone users.

1. Brain Exercise: This is a free app that provides a set of smart games to help you improve your Math skills starting from the basics.

2. Fallin Math: This is another great Math Game that offers different layers of difficulty to players. What is cool about this game is that users can easily share their scores with other with one click.

3. Math Workout: Math/Maths Workout is a set of daily brain training exercises and math drills designed to enhance mental arithmetic.

4. Mental Math Free: It gives useful tips to simplify difficult math problems and exercises to use them. It has four arithmetic operations training in three difficult levels.

5- Einstein Math Academy: Playing the game is easy (finding equation that works), but mastering it is hard due to the cleverly crafted scoring system. Make strategical decisions and improve skills to plan ahead.

6. Algebra Tutor: Walk through step-by-step solutions to see where you made your mistake. See your stats for every problem type (saved across app runs). Work out problems without needing to take notes.

## VI. IMPLEMENTATION AND ANALYSIS

To perform the implementation, we have selected two sections of students opted for learning discrete mathematics course in the College of Applied Sciences-Salalah. For one section we have used m-learning tools throughout the semester for teaching and learning and named that section as A. Another section which is named as B used normal e-learning tool (Blackboard) only. Section A students were having the advantage of accessing e-learning tool also in addition to the m-learning tools. We took much effort in the sample selection for this research so that the outcome would be accurate. We were concentrating on the equality in the samples selected. We had selected both groups from CAS-S for this course during fall semester where the average performance of the students was same in the pre-requisite course. We kept the usual materials in the Blackboard and gave links to the video lectures that was available in the YouTube through Blackboard itself. We were having same procedures to conduct the normal classes and exams. Same question papers were used to evaluate the student's performance in this course for both the sections. We asked one section students to install the hybrid e-book and math apps. Another section was not provided with these options. Using the statistical option given in Blackboard we were monitoring the number of students accessing the Blackboard during their semester in both the sections. Also using Student Information System (SIS) software's result options, we were comparing the results and dropout numbers of the students from both the sections of this course. This research was conducted for one semester. At the end of the semester the collected data were analyzed and produced in this section. Apart from the collected data, students were asked to provide feedback about the hybrid e-book, math apps and to give periodic report of the usage of e-book, math apps and its difficulties in using it. An academic team was employed in preparing e-book with required content and short video of the lectures and the exercise demonstration. We were very specific about making the content as short as possible to make it easy and comfortable for the students to read and watch. For all other interactive sessions we had taken the help of Blackboard.

The research team desired to know the impact of m-learning tools in the accessibility of the Blackboard. So the research team was focusing on the number of times the Blackboard was accessed for viewing the materials and also number of posts in the forum. The table 2 is clearly showing that the students of section A have not accessed the materials more than 2 times on an average throughout the semester. But the section B students have accessed the materials on an average of 8 times. The forum access was more from the section A students than section B. Section A were having more than 24 posts on an average by each student compared to section B students, who were having an average of 11 posts throughout the semester.

TABLE 2
ACCESSIBILITY OF BLACKBOARD BY SECTION A AND B STUDENTS

| | **Section A** | **Section B** |
|---|---|---|
| **Accessed Course Materials** | 2 | 8 |
| **Accessed Forum for discussions** | 24 | 11 |

In table 3 we have recorded the registration and result details of section A and B students. Section A students who received the m-learning tools did not withdraw the course in the middle, where in section B, 2 students withdrew the course in the middle. Similarly all the students passed the subject from section A, where 2 students failed the course in section B.

TABLE 3
STUDENTS REGISTRATION & RESULT DETAILS OF SECTION A AND B

| | Section A | Section B |
|---|---|---|
| **No. of students registered** | 27 | 25 |
| **No. of students withdrawn from course in the middle** | 0 | 2 |
| **No of students passed the course** | 27 | 21 |

From the above two tables we can easily find out that the section A students performance is better compared to section B. Since both the section students are having similar background and handled by the same research team, we strongly feel that m-learning tools motivates the section A students positively for better learning compared to section B students.

In the below given Figures 1 and 2, let us see the grade distribution of section A and section B students. These graphs again prove that the grades of section A students are better than section B.

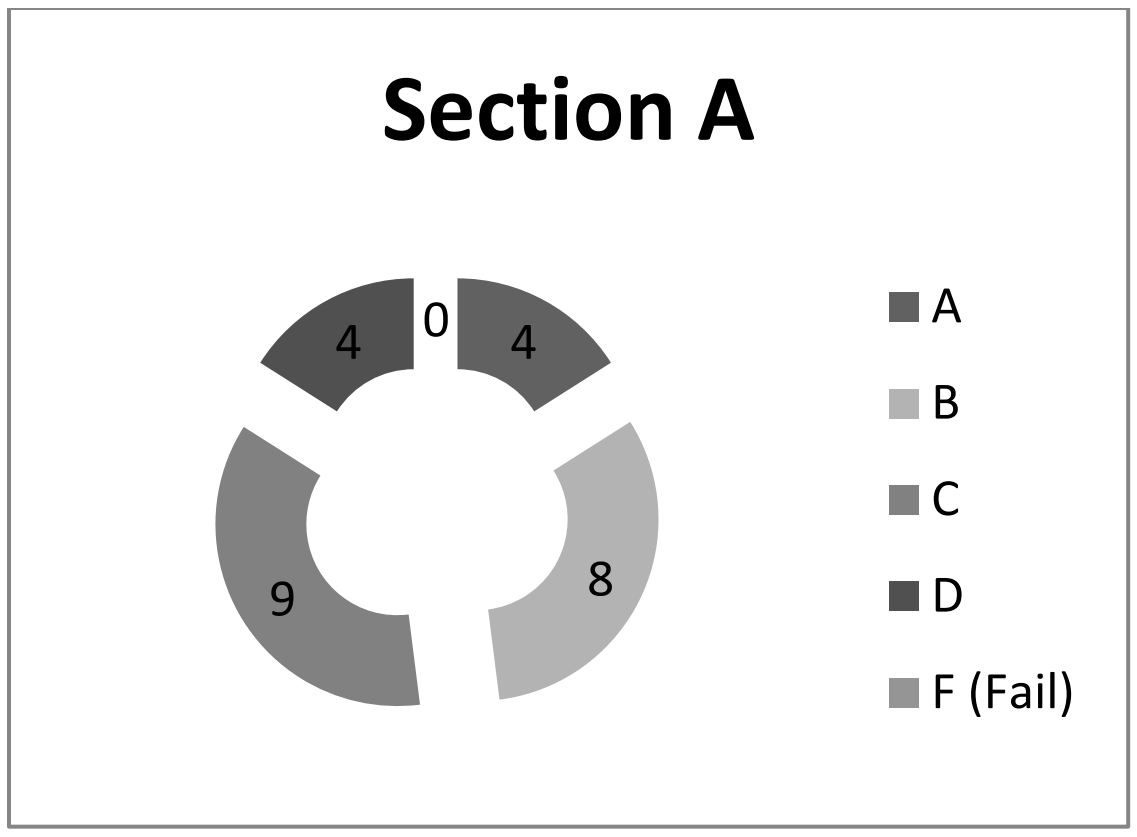

Fig 1: Grade distribution of Section A students

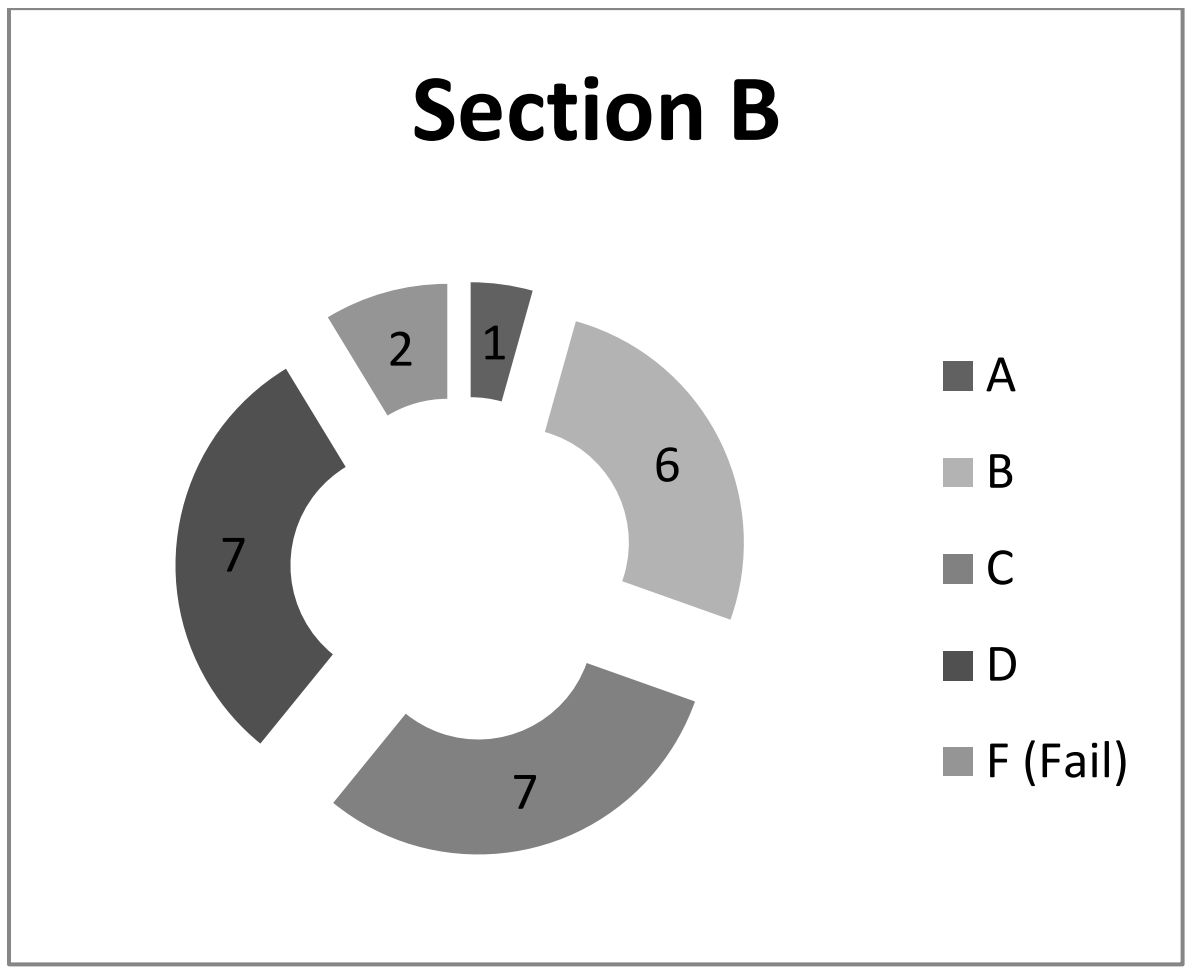

Fig 2: Grade distribution of Section B students

## VII. RECOMMENDATION

In the previous section we have discussed the advantages of m-learning in the higher education institutions in the Sultanate of Oman. Still we have found that the students are having few difficulties in using m-learning and expecting few more options for the maximum benefit from this. Most of the tools in m-learning are in English and few students are having difficulty in understanding the language, they want to have dictionary option in the tools of m-learning itself. In majority of the m-learning tools "Notes Taking" options are missing. Students are willing to have the notes taking option in their mother tongue or it should help the students to accept handwritten option. M-learning tools are having lots of advantages, but when a student wants to interact with teacher or wants to have access to the forum for that course, there is no facility provided with these tools. Even though there are separate chat options available in m-learning, incorporating chat option either in app or in the e-book will improve the convenience of interactive session between student and the peer. Similarly the students want to have the forum facility also to be accessed with the m-learning tools. Due to more multimedia facilities in the hybrid e-books, the accessibility of the books is comparatively slow in m-learning. Hence the speed of accessing the books should be more compared to the present accessibility speed. Since this research is conducted in the limited number of students in a higher education institution in Oman, we have got only few suggestions and could identify limited drawbacks with the m-learning. By expanding this research to other higher educational institutions we can improvise the m-learning to a great extent.

## VIII. CONCLUSION

The major problems with math course are that students were not able to link what they have learnt in these courses with the real world applications in their domain. Similarly math courses demand lots of practice and problem solving skills compared to other courses. Hence students need to practice the course materials even after their class hours. Now most of the world universities, colleges and other academic institutions are working towards teaching and learning these math courses which can make the students to access the study materials by JIT (Just-In-Time) and anywhere-anytime. This will help the students to access the materials at their potential time and reduces the knowledge loss. This paper discussed about the hybrid e-books, flipped classrooms and smart phone apps related to math to be implemented at the higher education institutions in the Sultanate of Oman and also proved that this methodology helps the students to perform well in their exams and other academic activities. This can be achieved by the students when they have hazel free method to access the materials relevant to the math course with additional support. M-learning is the excellent methodology to act as a platform for implementing contemporary tools of ICT like hybrid e-books, flipped classrooms and math learning apps. Even though M-learning is solving all the issues in the implementation of these technologies, still we keep the e-learning option open to the students who wish to use this sometimes for their comfort. This paper also proved that the students prefer M-learning than E-learning for the access of these contemporary ICT technologies for learning math courses. Similarly, m-learning helps students to get in touch with the materials as and when they wish, and improve their performance in exams. Still there are few disadvantages in the current version of m-learning tools for math courses which are pointed out in this paper's implementation section. By improving these options, in future we can make the math courses more interesting and likable even for the average learners.